# High Temperature Superconductivity in the Cuprates

B. Keimer[1], S. A. Kivelson[2], M. R. Norman[3], S. Uchida[4], J. Zaanen[5]

**The discovery of high temperature superconductivity in the cuprates in 1986 triggered a spectacular outpouring of creative and innovative scientific inquiry. Much has been learned over the ensuing 28 years about the novel forms of quantum matter that are exhibited in this strongly correlated electron system. This progress has been made possible by improvements in sample quality, coupled with the development and refinement of advanced experimental techniques. In part, avenues of inquiry have been motivated by theoretical developments, and in part new theoretical frameworks have been conceived to account for unanticipated experimental observations. An overall qualitative understanding of the nature of the superconducting state itself has been achieved, while profound unresolved issues have come into increasingly sharp focus concerning the astonishing complexity of the phase diagram, the unprecedented prominence of various forms of collective fluctuations, and the simplicity and insensitivity to material details of the "normal" state at elevated temperatures. New conceptual approaches, drawing from string theory, quantum information theory, and various numerically implemented approximate approaches to problems of strong correlations are being explored as ways to come to grips with this rich tableaux of interrelated phenomena.**

**Introduction**: The discovery of high temperature superconductivity in the cuprate perovskite LBCO[1] ranks among the major scientific events of the 20$^{th}$ century – it triggered developments in both theoretical and experimental physics that have significantly changed our understanding of condensed matter systems. Most obviously, the superconducting transition temperatures in the cuprates exceed those of any previously known superconductor by almost an order of magnitude, passing by a large factor what was, at the time, widely believed to be the highest possible temperature at which superconductivity could survive (Fig. 1). Moreover, according to all the intuitions developed based on the theory of "conventional" superconductors, the cuprates would have seemed the least likely materials in which to look for superconductivity at all: at room temperature, they are such poor conductors that they can hardly be classified as metals, and indeed if the chemical composition is altered very slightly they become highly insulating antiferromagnets. Magnetism arises from strong repulsive interactions between electrons, while conventional superconductivity arises from induced attractive interactions, making them seemingly antithetical forms of order.

As the properties of the cuprates were studied with ever increasing precision and sensitivity, it rapidly became clear that much of the well understood quantum theory of the electronic properties of solids, which has been spectacularly successful in accounting for the properties of conventional metals and superconductors, fails entirely to address many features of the cuprates, and more generally of a broad array of "highly correlated electron systems" of which the cuprates are the most studied example. A schematic phase diagram of the cuprates is shown in Fig. 2. Here, the x-axis is a material property, the "doping level", which controls the electron concentration per copper site in the all important Cu-O planes, and the y-axis is the temperature, T. The unprecedentedly high $T_c$ and the intimate relation between superconductivity and antiferromagnetism are only two of the unexpected features of this phase diagram. By way of introduction, we list some of the salient features which will be discussed below:

1) The origin and character of the antiferromagnetic state which is the "parent" of the high temperature superconductors is well understood from the strong-coupling perspective in which the insulating character derives from the classical repulsion between two electrons on the same atom, and the antiferromagnetism from the superexchange interaction, J, which is itself inversely proportional to U.[2,3]

2) Various attempts to obtain a semiquantitative estimate of the superconducting transition temperature, $T_c$, have had some measure of success[4], but there are important reasons, which we will discuss, to consider this problem still substantially unsolved. Certainly, there has not yet been any salient success in the theoretical prediction of new high temperature superconductors, or even in predicting which small changes to existing materials would produce increases (or decreases) in $T_c$. Recent developments in numerical methods for handling the physics of strongly interacting electrons from short to intermediate length scales[5], some of them informed by imports from quantum information theory, may provide an avenue for progress here, coupled with new ways of preparing and manipulating copper-oxide materials[6].

3) A conventional superconductor has the same symmetries as the underlying crystal, while in the cuprates it has "d-wave" symmetry[7,8], which means that the superconducting wavefunction changes sign upon rotation by 90°. Associated with this "unconventional pairing" is the existence of zero energy (gapless) quasiparticle excitations at the lowest temperatures, which make even the thermodynamic properties entirely distinct from those of conventional superconductors which are fully gapped. The reasons for this, and its relation to a proximate antiferromagnetic phase, are now well understood, and indeed were anticipated early on by some theories[9,10,11]

4) However, superconductors with unconventional symmetries were believed (and in some cases have since been shown to be) extremely fragile in the sense that $T_c$ is readily suppressed to zero by even small concentrations of impurities or crystalline imperfections[12], while superconductivity in the cuprates is quite robust to many forms of disorder. This has been suggested as being a consequence of strong local correlations[13], but still remains a puzzle.

5) The state at temperatures just above $T_c$, out of which the superconducting state condenses, is in most cases the "pseudogap" which is characterized by a substantial suppression of the electronic density of states at low energies that cannot be *simply* related to the occurrence of any form of broken symmetry. While much about this regime is still unclear, increasingly clear experimental evidence has recently emerged that there are strong and ubiquitous tendencies toward several sorts of order or incipient order, including various forms of charge density wave (CDW), spin density wave (SDW), and electron nematic order, and possibly pair-density wave (PDW) and orbital loop current (OLC) order, all of which compete with uniform d-wave superconductivity and exhibit similar energy scales. While there are many fascinating aspects of these "intertwined orders" that remain to be understood, their existence and general structure were in fact anticipated by theory[14,15]. Superconducting fluctuations also play a significant role in part of this regime, although to an extent that is still much debated.

6) The fact that at temperatures well above $T_c$, the conductivity is almost two orders of magnitude smaller than in simple metals and exhibits frequency and temperature dependences that are incompatible with the conventional theory of metals has led to this regime being referred to as a "strange metal" or "bad metal". The exhibited behavior, which is simple to describe in terms of the so-called "marginal Fermi liquid phenomenology"[16], has resisted any generally accepted understanding. On the other hand, similar behavior has now been documented in a large number of electronically interesting materials[17], indicating that this is a general property of strongly correlated electron systems, and not directly linked to high temperature superconductivity. We consider this to be the most significant open problem in the understanding of quantum materials, and it is here that radically new ideas, including those derived from recently developed non-perturbative studies in string theory, may be needed.

**Background:** The BCS theory[18] of the late 1950s provided an extremely successful framework for understanding conventional superconductors, and in the process gave rise to conceptual breakthroughs which affected the course of physics more broadly for decades to come: From it sprang an appreciation for the role of topology in physics (*e.g.* the Josephson effect[19] and vortex matter[20]), as well as deep notions of the roles of broken symmetry and the Anderson-Higgs phenomenon (the formalism that led to the prediction[21] of the recently observed Higgs boson). The basic insight is that the electrons collectively bind into "Cooper" pairs and simultaneously condense in much the same way as bosons condense into a superfluid state. Fundamental to the BCS mechanism is the fact that, despite the strong direct Coulomb repulsions, the relatively weak attractions between electrons induced by the coupling to the vibrations of the lattice (phonons) can bind the electrons into pairs at energies smaller than the typical phonon energy, $\hbar\omega_D$. However, this also implies that the superconducting $T_c$ is always relatively small, as it must satisfy a hierarchy of inequalities, $k_B T_c \ll \hbar\omega_D \ll E_F$, where $E_F$ is the Fermi energy. This bound was widely believed to imply that $T_c$ of conventional superconductors could never exceed 30K[22], although it has been revised upwards by the discovery in 2001 of superconductivity with $T_c$=39K in the simple metal $MgB_2$[23], where circumstances conspire to optimize the electron-phonon mechanism. However, this is still far below the maximal $T_c$ of the cuprates (Fig. 1).

The mystery was further deepened by the realization that the "chemistry" of the copper oxides amplifies the Coulomb repulsions. The all-important two dimensional copper oxide layers (Fig. 3) are separated by ionic, electronically inert buffer layers. The stoichiometric parent compound (Fig 2, zero doping) has an odd integer number of electrons per $CuO_2$ unit cell (Fig. 3) and is a "Mott insulator"[2]: In conventional band insulators, the electrons are effectively localized due to interference associated with the quantum mechanical wave properties of the electron, while in Mott insulators, electron motion is prevented by a classical jamming effect associated with the electron-electron repulsion. Specifically, the states formed in the $CuO_2$ unit cells are sufficiently well localized that, as would be the case in a collection of well-separated atoms, it costs a large energy (the Hubbard "U") to remove an electron from one site and add it to another. When there is an integer number of electrons per unit cell, the effect is quite literally like a traffic jam of electrons[24] where nothing can move. However, the electrons also carry spin and the zero-point quantum kinetic energy associated with localizing electrons produces an antiferromagnetic ("exchange") interaction between neighboring spins. This, in turn, leads to a Néel ordered phase below room temperature, in which there are static magnetic moments on the Cu sites with a direction that reverses from one Cu to the next[25].

The Cu-O planes are 'doped' by changing the chemical makeup of interleaved "charge-reservoir" layers so that electrons are removed ("hole doped") or added ("electron doped") to the copper-oxide planes (the horizontal axis of Fig. 2). In the interest of brevity, we will confine our discussion to hole doped systems. Hole doping rapidly suppresses Néel order. At a critical doping around $x_{min} \approx 5\%$, superconductivity sets in, with a transition temperature that grows to a maximum at about $x_{opt} \approx 16\%$, then declines for higher dopings and vanishes for $x_{max} \approx 27\%$. Materials with $x_{min} < x < x_{opt}$ and $x_{opt} < x < x_{max}$ are referred to, respectively, as "underdoped" and "overdoped".

The strong electron repulsions that cause the undoped system to be an insulator (with a 2 eV energy gap) are still the dominant microscopic interactions, even in optimally doped cuprate superconductors. This has several general consequences: i) The resulting electron fluid is "highly correlated" in the sense that for an electron to move through the crystal, other electrons must shift to get out of its way. This is at odds with the Fermi liquid description of simple metals, in which quasiparticles (which can be thought of as dressed electrons) propagate freely through an effective medium defined by the rest of the electrons. In some cases, though, an emergent Fermi liquid arises at low temperatures. This is especially clear in the overdoped regime (Fig. 2), but even in underdoped materials, at low enough temperatures when superconductivity is quenched by the application of a high magnetic field, emergent Fermi liquid behavior is also observed, albeit with characteristics (*e.g.* a reconstructed Fermi surface) that are quite different from those predicted by the simplest band theory[26]. Still, over most of the phase diagram, the frustration of the coherent electron motion produces physics that is qualitatively distinct from that of simple metals. ii) While the large zero-point energy of electrons in a usual metal results in a quantum "rigidity" which greatly suppresses all forms of inhomogeneous states, the Mott physics and the short-range antiferromagnetic correlations inherited from the undoped "parent" compound combine to produce a local tendency to phase separation and various forms of order which spontaneously break the translational symmetry of the underlying crystal[27,58,56]. Thus, especially in the pseudogap regime of the phase diagram, various forms of nematic, CDW, and SDW order occur on intermediate length scales. iii) The failure of the quasiparticle paradigm is most acute in the "strange metal" regime that is the "normal" state out of which the pseudogap and the superconducting phases emerge when the temperature is lowered.

In the remainder, we address the various regimes of Fig. 2 in more detail, finalizing our discussion with a sketch of recent theoretical developments. But before doing so, let us first discuss in detail why a high superconducting temperature in and of itself no longer appears to be an essential part of the mystery.

**Pairing by repulsive interactions: the unconventional superconductor**

It is now well established that electrons can form pairs, even when they repel each other at a microscopic scale. However, this involves non-trivial physics. A model that is often used as a point of departure for theoretical discussions is the famous Hubbard model, describing electrons hopping on a lattice parametrized in terms of the bandwidth $W = 8t$ (where t is a measure of the 'hopping' energy gain due to delocalization of the electrons) and an on-site electron-electron repulsion U. It is generally accepted that in the cuprates, U and W are comparable. Even for this simplified model, analytic solutions are not available, let alone for more realistic Hamiltonians which better capture the local solid-state chemistry. However, approximate solutions of the doped Hubbard model can be obtained in several ways, and these invariably point to a d-wave superconducting ground state, something that had already been found in theoretical studies completed just before the discovery of superconductivity in the cuprates[9,10,11].

An intuitive understanding of the mechanism of pairing is best obtained by approaching the problem from an unrealistic weak-coupling perspective, *i.e.* assuming $U \ll W$[28]. Here, the gap structure is determined by the solution of a variant of the original BCS equations in which an appropriately renormalized two-particle vertex function, $\Gamma(\mathbf{k})$, plays the role of an effective interaction. For the case of purely repulsive interactions, if $\Gamma$ is sufficiently $\mathbf{k}$ dependent, a sign-changing superconducting order parameter (where $\Delta(\mathbf{k})$ and $\Delta(\mathbf{k+Q})$ have opposite sign) results where interactions involving small momentum transfer are pair-breaking, while those with large momentum transfer near $\mathbf{Q}$ promote pairing. In particular, if there are antiferromagnetic correlations, this typically implies a peak in $\Gamma$ at the antiferromagnetic ordering vector, $\mathbf{Q}=\mathbf{Q}_{AF}$[29], which is also an ideal vector for scattering between 'antinodal' regions of the Fermi surface of the cuprates shown in Fig. 4, *i.e.* precisely those regions where the d-wave gap is largest and of opposite sign. The gap 'nodes' along the diagonals of the Brillouin zone are then, in turn, where the d-wave gap vanishes.

Superconductivity in the Hubbard model cannot truly be approached from the strong-coupling limit, since there is now strong numerical evidence that for a broad range of doping, the ground state of the Hubbard model is ferromagnetic rather than superconducting for large enough $U/t$[30]. However, the closely related t-J model (with the superexchange $J=4t^2/U$) incorporates the essence of the strong coupling physics through the constraint that no more than one electron at a time can occupy a given site. The t-J model can then be addressed with values of $J/t \sim$ ½ as a reasonable model in its own right. While no controlled solution is known, the superconducting tendencies of this model have been investigated since the early days of high temperature superconductivity using variational projected (RVB) wavefunctions and slave-particle mean-field theories[31,32]. It is striking that the character and symmetry of the superconducting state itself and its association with short-range antiferromagnetic correlations look grossly similar, regardless of perspective[29].

While intermediate coupling problems have thus far not been successfully solved by controlled analytic approaches, the lack of any small dimensionless parameters likely implies the lack of any long emergent length scales in the problem. With this in mind, a variety of numerical techniques have been spawned to study this regime, including exact diagonalization (limited to small clusters), quantum Monte Carlo and its derivatives (variational and fixed node approximations to get around the issue of negative probabilities in fermion simulations), dynamical mean field theory[33] (DMFT) and its cluster generalizations (either in momentum space or real space), density matrix renormalization group (designed for 1D problems but can simulate strips), and its 2D generalizations (PEPS - projected entangled pair states[36] and MERA - multi-scale entanglement renormalization[34]). These methods all have their pluses and minuses, in particular in regards to bias. They have, however, taught us that if superconductivity occurs, it is invariably of d-wave symmetry, but also that many competing states are close by in energy, especially unidirectional charge order[5,35,36].

**Why the high $T_c$ superconductor is even more different**

How do these theoretical results relate to the real experimental systems? Static antiferromagnetism disappears quickly as a function of doping (Fig. 2), but both inelastic neutron scattering (INS) and resonant inelastic X-ray scattering (RIXS) reveal that the antiferromagnetism of the insulator survives in the superconductor to a degree in the form of *dynamical* magnetic fluctuations which are much stronger than in conventional metals (and are strongly renormalized when cooling below $T_c$)[37,38,69,70]. The simplest thing to do is to just take this measured spin fluctuation spectrum and construct the vertex Γ mentioned above. This appears to yield reasonable values for $T_c$, while it also seems consistent with some of the single electron self-energy effects detected by various electron spectroscopies such as ARPES (angle resolved photoemission) and STS (scanning tunneling)[4]. In various ways, though, this "spin fluctuation glue" approach has shortcomings[39]. Despite its intuitive appeal, it is not based on controlled mathematics, since the same electrons one is pairing are also forming the "glue". Another difficulty is that these simplified models leave out other effects that can influence the magnitude of $T_c$. A case in point is the electron-phonon interaction. There is significant evidence that phonons affect both ARPES and STS lineshapes[40], while strong anomalies are seen in the phonon spectra[41]. There are a number of other neglected effects that are worrisome, in particular the non-local Coulomb interaction which is an especially relevant concern given the poor screening in the direction perpendicular to the planes[42]. Interplanar effects are also known to play an important role in the physics of $T_c$. In homologous series containing multiple $CuO_2$ planes, n, $T_c$ typically increases up to n=3, and then decreases for higher n[43]. Moreover $T_c$ is sensitive to the location of the off-planar (apical) oxygens that are located above the Cu ions[44].

More seriously, some fundamental aspects of high $T_c$ superconductivity are *qualitatively* different from the BCS variety. An example is the influence of quenched disorder. In absolute terms, all cuprates can be regarded to be chemically quite "dirty" due to their doped nature. In BCS theory, a great contrast is found between s-wave and unconventional superconductors, where any form of potential disorder should be detrimental for the latter[12]. The strong inhomogeneity seen by STS leaves no doubt that many cuprate superconductors live in a disordered lattice potential, but the d-wave superconductor is fairly insensitive to this adverse condition. However, the disorder associated with substitution of Zn or Ni ions for the Cu ions does exert a detrimental effect as expected for a d-wave superconductor[45].

A very basic quantity for the superconducting order is the superfluid density $\rho_s$, the quantity that parametrizes the rigidity of the phase of the superconducting order parameter, which also determines the capacity of the superconductor to expel electromagnetic fields. One can identify a temperature associated with the fluctuations of the phase $T_\theta \sim \rho_s/m^*$ where $m^*$ is the effective mass. In a BCS superconductor, $\rho_s$ is equal to the total density of electrons at zero temperature and accordingly $T_\theta \gg T_c \sim \frac{1}{2} \Delta_0/k_B$, the temperature associated with the formation of the pairs where $\Delta_0$ is the average superconducting gap. The fluctuations of the phase of the superconducting condensate are largely irrelevant; once Cooper pairs form, they automatically condense. Turning to the cuprates, it was established early on that the superfluid density is anomalously small, scaling in the underdoped regime with $T_c$ (the "Uemura[46] law"). The conclusion is that in the underdoped cuprates, $T_\theta$ and the pair binding energy are of the same order and the thermal fluctuations of the phase should be crucial for the thermodynamics[47]. A long standing question is whether perhaps pairs already form at the (very high) pseudogap temperature T* (Fig. 2) while at a much lower temperature, the actual $T_c$, the phase locks to form the long range ordered superconducting state. As we will see, the physics of the phase fluctuations is intertwined with that of competing order.

The best superconductors are found at (or near) optimal doping, where one is dealing with an instability of the "strange metal" phase. The strange metal is the least understood part of the phase diagram, as it does not appear to be describable in terms of Landau quasiparticles. The very non-BCS transition from the physics of the strange metal to the more conventional physics of the superconducting state is vividly apparent in the temperature evolution of the ARPES spectra at momenta near the "antinodes" (Fig. 4) where the pairing forces of the d-wave superconductor are supposedly the strongest. For these momenta, the electron spectral function is strongly broadened as a function of energy[48]. Upon entering the superconducting phase, a quasiparticle peak starts to develop which has the classic Bogoliubov dispersion of a BCS superconductor[49]. This is in turn consistent with the onset of coherence in microwave, infrared and thermal conductivity. But, unlike in BCS theory, it appears that the spectral weight of this antinodal "Bogoliubon" is linearly proportional to the superfluid density, both as a function of doping and temperature[50,51]. This is not understood: it is as if the phase coherence of the superconductor "freezes out" quantum mechanical coherence from the highly collective non-Fermi liquid strange metal state. In the end, the understanding of the superconducting ground state, and why $T_c$ is so high, might well be buried in the presently enigmatic physics associated with the strange metal.

**Pseudogap regime and the competing orders**

The physics in the pseudogap regime has been the focus of much recent research. What has emerged is that a variety of orders different from superconductivity are at play here. Some involve "crystallization" of the electrons, in the form of stripes and other forms of charge order, but others appear to be more novel quantum liquids. There are signs of a fluctuating form of order that resembles spin singlets. If true, this would mean that local pairs might already be formed at very high temperatures. There is also some evidence for a new type of order involving orbital currents.

In the underdoped regime, the doped holes remain mobile even at low temperatures, but organize themselves collectively. A prominent feature of this regime of the phase diagram is the line T* which denotes the onset of a partial gap observed in spectroscopic data. First inferred from NMR measurements which showed a reduction in the low-frequency spin excitations[52], this pseudogap was subsequently seen in c-axis polarized infrared conductivity measurements and is associated with a pronounced upturn in the c-axis resistivity[53]. In contrast, the in-plane polarized infrared conductivity indicates a drop in the scattering rate[54], which is reflected in a reduction of the planar resistivity[55].

The idea that this could be related to forms of order other than pairing was brought to light in the mid 1990s by the experimental discovery of electronic "stripes" in the LSCO family[56]. This was inspired by earlier theoretical work that discovered that the mean field solution for doped Mott insulators on a square lattice in the intermediate coupling regime consists of Mott-insulating antiferromagnetic domains, separated by a regular "stripe" array of antiferromagnetic domain walls trapping the doped holes[27]. These were predicted to be insulators, and have now been seen in non-cuprate doped Mott-insulators that are characterized by larger values of the spin S and stronger electron-phonon interactions (manganites, cobaltates, and nickelates)[57]. An alternative and complementary view is that the system of doped holes and Cu spins tend to phase separate, but this is frustrated by the long range Coulomb interaction, and the compromise is to form conducting stripe-like textures[58].

In 1995 it was discovered by neutron scattering that such stripe-like order, characterized by incommensurate antiferromagnetic order and charge segregation, does occur in underdoped versions of LSCO where an "LTT" lattice deformation apparently acts as a pinning potential for the stripes[56]. However, it became clear that these stripes were different from the "classical" stripes in the other doped Mott-insulators: the cuprate stripes stay metallic and even superconduct at low temperatures. Although the spatial organization looks similar to the mean field stripes, a crucial difference is that these now can be viewed as a partially crystallized superconductor, formed from electron pairs.[36]

In general terms, a competition between superconductivity and crystallization is a very natural way to enhance the phase fluctuations of the former, yielding a rationale for the diminishing superfluid density in the pseudogap regime. Quite recently, evidence has emerged that materials with static stripes form a pair density wave: the charge stripes are internally superconducting, with a phase that reverses from stripe to stripe[14]. Given that the stripe orientation changes as one moves from one layer to the next, this frustrates the Josephson coupling between layers, giving rise to a two dimensional superconducting state consistent with the Kosterlitz-Thouless behavior seen in transport measurements[59,60].

It was subsequently found by INS that the spin-wave spectrum of the magnetically ordered stripes has a unique "hour-glass" pattern[61], with the neck of the hourglass located at the commensurate antiferromagnetic wavevector, and this pattern has subsequently been observed in the insulating charge ordered states of manganites and cobaltates[57]. In the cuprates, this pattern persists for larger dopings where the stripe order is no longer static, and is strongly modified below $T_c$[62]. A similar pattern is predicted for interacting fermions in a homogeneous d-wave superconductor[63]. A reconciliation of these two very different pictures remains a challenge for the field.

Static stripe order had seemed to be confined to the LSCO family. However, recently, incommensurate charge ordering was discovered in underdoped YBCO, BSCCO, and HBCO. X-ray experiments find short-range charge order that gradually onsets between 100 and 200K[64,65]. Moreover, high energy x-ray scattering[66] and NMR experiments have confirmed that the short-range charge order is truly static, and thus presumably arises from pinning of correlated charge fluctuations by defects[67]. A difference with the stripes in the LSCO family is that there is no evidence of coincident static (or nearly static) magnetic order. Moreover, the variation of the stripe wavevector with doping in YBCO is opposite that in LSCO[68]. Whereas in the latter, this wavevector increases with doping as expected in a real space picture, in the former, the wavevector decreases as would be expected from a momentum space picture involving the underlying Fermi surface. This difference may be connected with differences in the spin behavior: In YBCO a large spin gap is present which acts to suppress the incommensurate spin order that is more prevalent in the LSCO family[69,70].

In a parallel development, the structure of the pseudogap in momentum space was directly mapped by ARPES in 1996, showing that this to a degree mimics the d-wave superconducting gap: a gap was apparent only in the "antinodal" regions of the Brillouin zone (Fig. 4)[71,72,73]. The suggestion was immediate that already at the very high pseudogap temperature T*, pairs start to form while phase fluctuations prohibit superconducting order until much lower temperatures. So long as there is substantial short-range phase coherence, superconducting fluctuations should have large and identifiable signatures. For instance, for a range of temperatures that extends up to about 1.5 $T_c$ but *not* to temperatures comparable to T*, large fluctuation conductivity (both DC and AC) is observed[74], but there is some debate how much such signatures differ from those observed in classic superconductors. Moreover, the reconciliation of this superconducting-like signature in the fermion response and an energy gap due to crystallization has been a major challenge as well, even more relevant now given the new findings of CDW order mentioned above.

The more far-reaching notion of pairing correlations without substantial phase coherence persisting to temperatures of order T* is difficult to define precisely, even in principle. The best circumstantial evidences come from diamagnetism that is observed up to about 150K[75]. Though weak compared to full Meissner screening, it is still large compared to that of simple metals. Moreover, in underdoped YBCO, a moderately well defined interlayer Josephson plasma resonance seems to persist up to similar temperatures[76], and recent pump-probe experiments are consistent with transient superconducting order existing all the way to T*[77]. Perhaps the best evidence is in the temperature evolution of the gap itself. Despite the sense of "two gaps" as we will discuss next, the high temperature pseudogap evolves remarkably smoothly into the gap structure of the superconducting state.

Scanning tunneling spectroscopy has proven to be particularly revealing in this context. Such data (mostly below $T_c$) exhibit electronic waves in real space that upon Fourier transformation show peaks that disperse with bias and have been mapped to scattering across the Fermi surface[78]. One finds that in the superconducting state, the low energy excitations near the nodes behave just as one would expect for a BCS d-wave state[79], but at higher energies, cross over to a dispersionless pattern characteristic of short-range stripe order. Interestingly, this low energy dispersing pattern maps out a Fermi arc[80] as observed directly by ARPES[81] in the pseudogap state, with the arc recovering the full Fermi surface once the doping exceeds a critical value[82]. This is in contrast with ARPES that sees antinodal quasiparticles below $T_c$, even for underdoped materials[83]. The higher energy dispersionless pattern is seen at all energies when moving above $T_c$ into the pseudogap state[84,85], consistent with local charge order, and has been identified as co-existing with the low-energy QPI signal below $T_c$ as well[86,87]. But the consistency of the ARPES data with charge order is still an active area of debate – to date, no unambiguous signatures associated with the stripe wavevectors have been found.

Yet another interesting hint regarding the unusual relationship between the charge order and superconductivity follows from the temperature evolution of the charge order. The X-ray signal begins to build up smoothly upon cooling below some $T_{cdw}$ typically less than T* to attain a maximum at the superconducting $T_c$, then drops significantly below $T_c$, indicating competition between the CDW order and superconductivity[64,65].

There is also evidence for "quantum liquid crystal" order occurring in the pseudogap phase. Such phases are translationally invariant while they break the rotational symmetry. First suggested in the context of the quantum melting of stripe crystals[88], evidence appeared for such a phase breaking the fourfold symmetry of the square lattice in underdoped YBCO from transport[89,90] and INS measurements[91]. This "nematic" signal was also found in the analysis of the STS data, showing that besides the "stripy" texture breaking translations, there is also an overall (zero wavevector) breaking of rotations present, consistent with the two oxygen ions in the $CuO_2$ unit becoming inequivalent[92,93].

These orders are all close siblings of the electron crystal. However, there is also evidence for a completely different kind of order that sets in at T* itself. This order is symmetry wise equivalent to having magnetic moments on the oxygen sites, and thus would be a magnetic analogue of the charge nematic mentioned above[94]. But the original proposal that motivated the experiments involved spontaneous electron currents flowing inside the $CuO_2$ units in such a way that although rotational symmetry is broken, translational symmetry is not[95]. It has a quite distinct magnetic diffraction pattern that was subsequently seen by spin-flip neutron scattering[94]. Fluctuations associated with this order act like spin fluctuations in mediating d-wave pairing[96], but it does not yield a natural explanation for the pseudogap, just by the very fact that it does not break translational symmetry. The real difficulty with this proposal is that this current order should also be seen by local magnetic probes like μSR and NMR, but this has not been observed. Potentially related to this is the onset of a small Kerr rotation at T* which also indicates some type of symmetry breaking[97]. This Kerr signal defines a phase line that cuts through the superconducting dome, vanishing near 18% doping.

A much debated question is whether conventional (Hartree-Fock) mean-field treatments are able to provide even a qualitatively correct account of the pseudogap phenomenology. There are surely issues related to the "plethora of orders", something which is not easy to understand in this way. However, one obtains a much sharper view utilizing electron spectroscopies. The striking difference in the nature of the electronic excitations measured in ARPES when crossing from the "coherent" nodal region to the "incoherent" antinodal region in momentum space is called the nodal-antinodal dichotomy[98]. The nodal region involves a narrow region around the zone diagonals which gradually grows with increasing doping until it encompasses the entire Fermi surface in sufficiently overdoped materials. While the antinodal region lacks any quasiparticle-like peaks in the spectral function, throughout the pseudogap regime it exhibits a suppression of low energy spectral weight on an energy scale that corresponds to the pseudogap[48].

The astonishing character of these observations is best illustrated by showing a map of the spectral weight at low energy as a function of **k** in the first Brillouin zone (Fig. 4). In a Fermi liquid, the Fermi surface delineates the boundary between occupied and unoccupied quasiparticle states, so no matter how complicated it may be, the one thing it cannot do is abruptly end. However, in the pseudogap regime, there appear to be "Fermi arcs" in the nodal regime[81]. In a mean-field theory, the effective potential associated with a (density-wave) state that breaks translational symmetry can reconstruct a large Fermi surface, producing small Fermi surface pockets, but these still have to form connected manifolds. It is plausible that the Fermi arcs are actually the front half of such a pocket[71], and hence there has been an intense search to find the "backside of the pocket", but at present there is no definitive sign of it. As a function of decreasing temperature, a BCS-like gap opens on the arc, eventually merging with the antinodal gap into an overall gap structure that, at low temperatures, is not all that different from a simple BCS d-wave gap. Very recently, evidence has emerged that the momenta of the endpoints of the arc are directly related to the ordering wave vector of the charge order[99], though the precise definition of these arc tips is still a subject of active debate. In support of this, quantum oscillation studies are consistent with the arcs linking together to form an electron pocket centered at (Q/2,Q/2), where (Q,0) is the CDW wavevector[100]. This demonstrates that the arcs and the charge order are somehow related, but in a way that is completely different from arguments based on nesting, which would predict that this vector spans the antinodes instead[63]. And away from the ordered states, things become even more mysterious – the strange metal.

**The strange metal: quantum criticality versus the conformal metal**

The strange metal was early on recognized as perhaps the most mysterious aspect of cuprates. The gross difference of the strange metal phase from that of conventional metals is the absence of quasiparticles. This has consequences for simple physical properties like the electrical resistivity. In a normal metal, the resistivity saturates at high temperatures when the mean free path becomes of order the electron de Broglie wavelength. The resistivity of the cuprate strange metal can be linear in T from near $T_c$ up to as high a temperature as measured[101]. Moreover, the Hall resistivity has a different temperature dependence than would be expected in a quasiparticle picture[102].

In the late 1980s, these and various other experimental anomalies were encapsulated in the phenomenological "marginal Fermi liquid" theory[16]. This asserts that the Fermi gas is coupled to a continuum of excitations that is spatially featureless, with a spectral density which is constant for $\omega > T$, but proportional to T for $\omega < T$. This leads to a damping rate that scales as max($\omega$,T). This was confirmed later by high resolution ARPES measurements, with the caveat that this is only seen in the nodal region, with the antinodal region behaving in a more incoherent fashion[103].

In the 1990s the idea of quantum criticality emerged to explain the low energy excitations of the strange metal. A quantum phase transition occurs when a continuous phase transition occurs at zero temperature as a function of a tuning parameter (like pressure or doping), where the corresponding quantum critical point (QCP) defines the boundary between the ordered (broken symmetry) and disordered quantum phases[104]. The correlations at a QCP are characterized by a spatio-temporal scale invariance, which in turn has the effect that there are no longer quasiparticle poles in spectral functions. Instead one finds power law behavior ("branch cuts") and spectral functions at finite temperature that are scaling functions of $\omega/T$. This can be interpreted in terms of a dissipative energy relaxation time $\hbar/k_BT$, which is sometimes referred to as "Planckian dissipation" as it is a quantum effect independent of material parameters[105]. Moving away from the critical point, the energy scale above which scale invariance remains gradually increases. Accordingly, in the "tuning parameter"-temperature plane, there is a quantum critical wedge opening up from the QCP. This renders a suggestive interpretation of the phase diagram in Fig. 2, where the strange metal is identified with the quantum critical wedge associated with a QCP under the superconducting dome near optimal doping.

The theory of quantum criticality in metallic systems is still a work in progress. One issue is that there may be reasons to believe that the QCP is intrinsically unstable, since the order parameter fluctuations mediate attractive interactions that promote superconductivity, meaning that the quantum critical point might always be "shielded" by a superconducting dome, just as in Fig. 2. However, there is also typically a diverging correlation length at a QCP, while no such growing correlation length has yet been observed in cuprates for any of the orders that are considered likely candidates. Moreover, according to the marginal Fermi liquid phenomenology[16], what is needed is a rather special sort of quantum criticality that is local in space, and so featureless in **k**.

Is there a quantum critical point involving the termination of pseudogap order inside the superconducting dome? There is evidence for this to be the case, early on from specific heat data[106] and more recently from a Fermi velocity anomaly seen in photoemission[107] as well as from the Kerr rotation[97], with the latest being a divergence in the effective mass seen in quantum oscillation studies[108]. But *which* order parameter rules the quantum critical regime, and is that regime large enough to encompass the entire strange metal region? We argued in the previous section that the pseudogap is characterized by several competing ordering tendencies. Even more seriously, this quantum critical description should break down at higher (ultraviolet) temperatures. How to explain then that the resistivity stays linear in T up to temperatures where the crystal melts?

The deep issue is that for a highly correlated fluid, the interactions are large and so likely cannot be treated using any fundamentally perturbative approach which starts with a free particle description. There is a well developed and extremely successful theoretical solution of this problem applicable to 1D and quasi-1D electron fluids based on "bosonization", but no such approach exists in higher dimensions. In this context, it is important to seek new approaches, theories that honestly treat the strong correlation physics, even if the connection to the relevant microscopic physics is unclear. This is where the mathematics of string theory might help: with the so-called holographic duality, one can address the physics of strongly interacting finite density systems. The working horse is a mathematical edifice called the AdS/CFT correspondence that has become a central subject in modern string theoretical research. Discovered in 1997[109], it demonstrates that the two grand theories of physics that seem unrelated (general relativity and quantum field theory) can become under certain conditions two sides of the same coin. According to the correspondence, there is a "holographic" relation in the sense that the quantum field theory is like a two dimensional photographic plate with interference fringes encoding the gravitational physics in three dimensions. Most importantly, the difficult to solve strong coupling quantum field theory is mapped to its more easily solved "dual", the weak coupling gravity theory.

Since 2007, the properties of matter at finite density have been the central focus of this "holography" research[110]. At low temperatures one finds superconductors, stripe and current phases, and even Fermi liquids. The observable responses of these states are often similar to experimental observations. However, the great difference is in the nature of the strange metal at higher temperatures. The gravity "dual" tells us that these systems at finite fermion density should form metallic quantum critical phases, where the scale invariance emerges *without fine tuning to any special quantum critical point*. However, these "conformal" metals (that exhibit Planckian dissipation) are intrinsically unstable and upon cooling spawn an extensive manifold of stable states. They also have special scaling properties that are different from any conventional quantum critical state. A case in point is that the simplest gravitational solution (an "extremal" charged black hole) translates into a conformal metal showing *local* quantum criticality, meaning that the metal is temporally quantum critical but spatially normal[111].

Due to the limitations of the mathematics, holography can only be proven in certain field theories that have no resemblance whatsoever with the electrons in the cuprates. It cannot be decided whether the traits discussed above are ubiquitous emergent phenomena or somehow tied to these special cases. At the least, however, holography can supply powerful metaphors, teaching physicists to think differently, leading to new questions to ask in experiments.

**The overdoped regime: back to normal?**

As the doping is increased beyond optimal, it appears that a real Fermi liquid begins to be established: quantum oscillations indicate a well developed large Fermi surface, consistent in detail with the prediction of one electron band theory[112]. This is supported by ARPES measurements where now sharp peaks are observed near this Fermi surface throughout the Brillouin zone (including the antinodes)[48]. INS data indicate a dramatic suppression of magnetic spectral weight near the antiferromagnetic wave vectors, which may be alternatively interpreted as a disappearance of the spin-fluctuation pairing glue, explaining why $T_c$ goes down[113] (though recent RIXS data have demonstrated pronounced spin fluctuations at smaller wave vectors, implying that strong electron correlations persist even in highly overdoped cuprates[38,114]). A big question is how really different the Fermi liquid at lower temperatures is from the anomalous strange metal at higher temperatures. ARPES shows there is only a weak crossover line that separates these two regimes[115].

**Outlook**

Originally inspired by the desire to find out why superconductivity can happen at a high temperature, condensed matter experimentalists engaged in a resilient effort to unravel the physics of cuprates. They profited from spectacular progress in experimental techniques, especially spectroscopies, but also in the challenging work of preparing better crystals. Over the last quarter century, this has turned into a serendipitous scientific endeavor that can be taken as a role model for other fields in physics. At stake is the general nature of quantum matter, and we have tried to explain in this overview that the established understanding of this matter falls short in several respects. As we also emphasized there is still plenty of work to do, especially with regards to the physics of competing order in the underdoped regime. However, in order to really penetrate these worlds of strongly interacting quantum systems, one needs the power of mathematics. This has become increasingly a bottleneck: although the theorists have been instrumental in guiding the experimentalists to discover various competing orders and introducing the ideas of quantum criticality and so forth, the atmosphere is distinctly different from e.g. cosmology or high energy physics where thousands of experimentalists invest many years in confirming strong theoretical predictions, like the Higgs boson or the B-modes in the cosmic microwave background. The bottom line is that the existing mathematical machinery seems inadequate in both describing the rich physics of the pseudogap phase, as well as the nature of the strange metal phase.

But times are changing. Experimental techniques to control correlated electrons are evolving rapidly and have the potential to lift the dialogue with theorists onto an entirely new stage. Recent examples include the development of atomically precise layer deposition methods that allow tailoring of lattice structures[6] and coherent optical control techniques[77]. In another serendipitous development, the builders of quantum information and string theory have landed in the same territory, finding out to their surprise that they are to a degree speaking the same language as condensed matter physicists. This is also reflected in the fact that some of the key underlying physics has been captured by advanced numerical techniques like DMRG and its descendants, which were also motivated by quantum information theory. The jury is still out on whether this is a coincidence or signals the onset of a revolution in physics.

———————

Affiliations:

[1]Max-Planck-Institut fur Festkorperforschung, Heisenbergstr. 1, D-70569 Stuttgart, Germany

[2]Department of Physics, Stanford University, Stanford, California 94305, USA

[3]Materials Science Division, Argonne National Laboratory, Argonne, IL 60439, USA

[4]Department of Physics, University of Tokyo, Bunkyo-ku, Tokyo 113-0033, Japan

[5]Instituut-Lorentz for Theoretical Physics, Universiteit Leiden, P.O. Box 9506, 2300 RA Leiden, The Netherlands

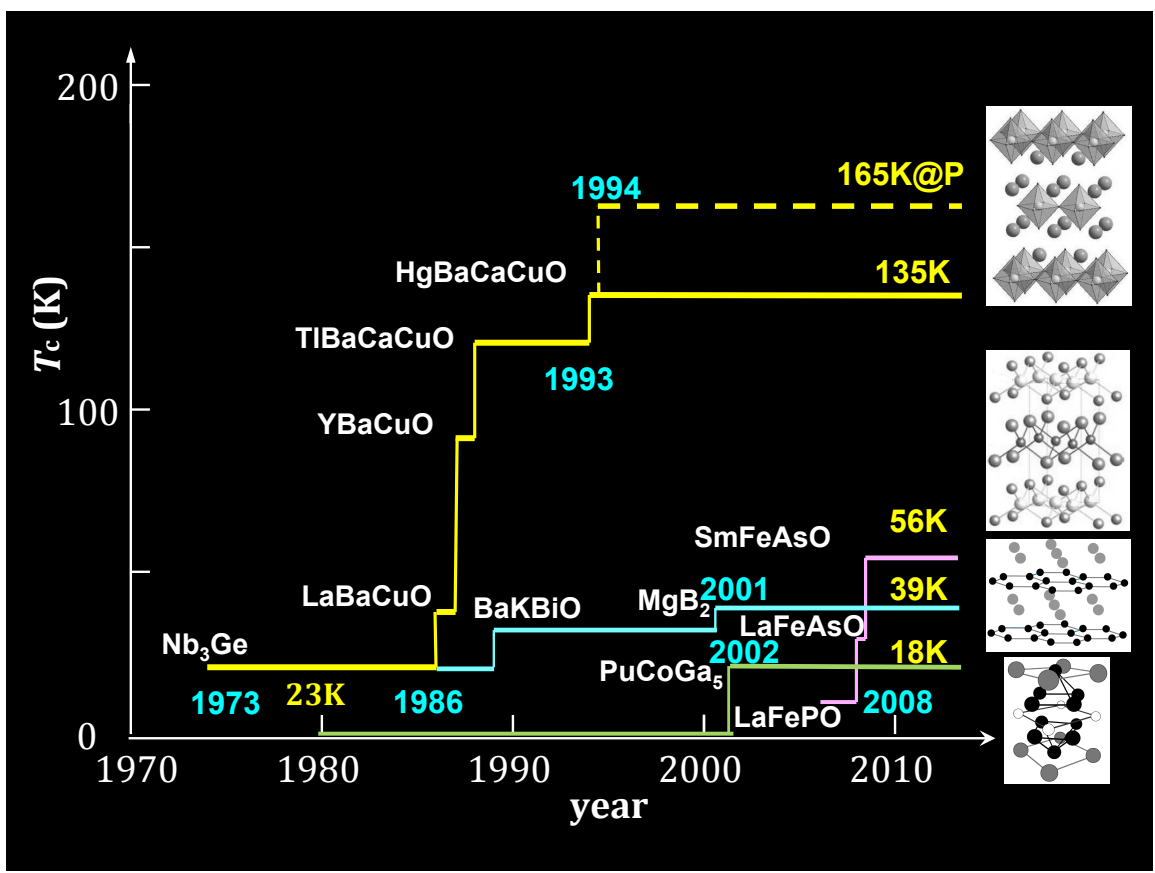

Figure 1 – **$T_c$ versus time.** Superconducting transition temperatures versus year of discovery for various classes of superconductors. The images on the right are the crystal structures of representative materials.

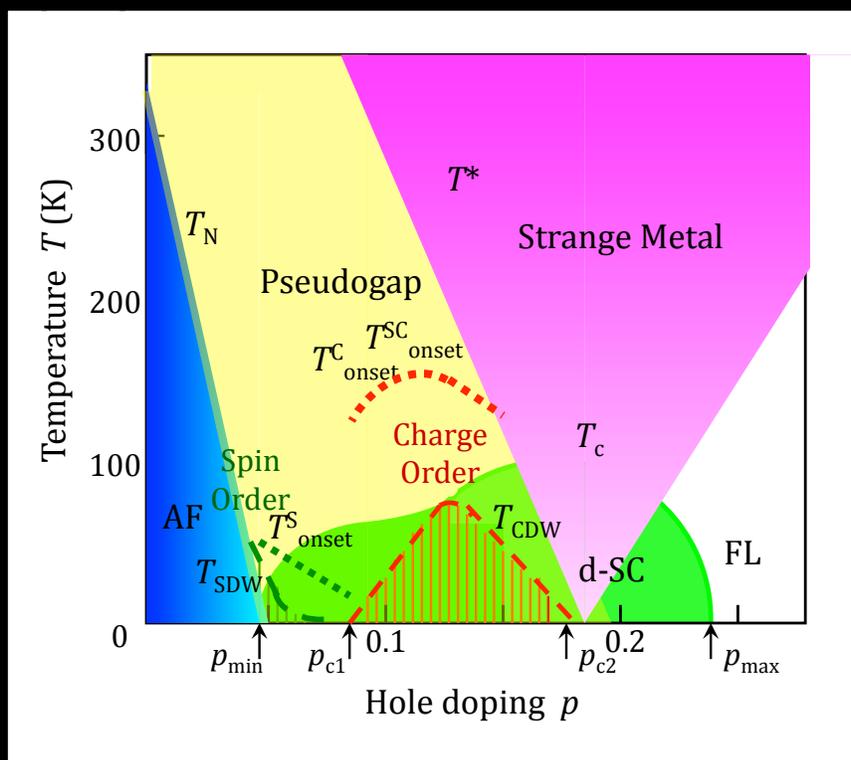

Figure 2 – **Phase diagram.** Temperature versus hole doping level for cuprates, indicating where various phases occur. AF is antiferromagnet, d-SC d-wave superconductivity, and FL Fermi liquid. SDW and CDW represent incommensurate spin density wave and charge density wave order. "onset" marks where precursor order or fluctuations become apparent.

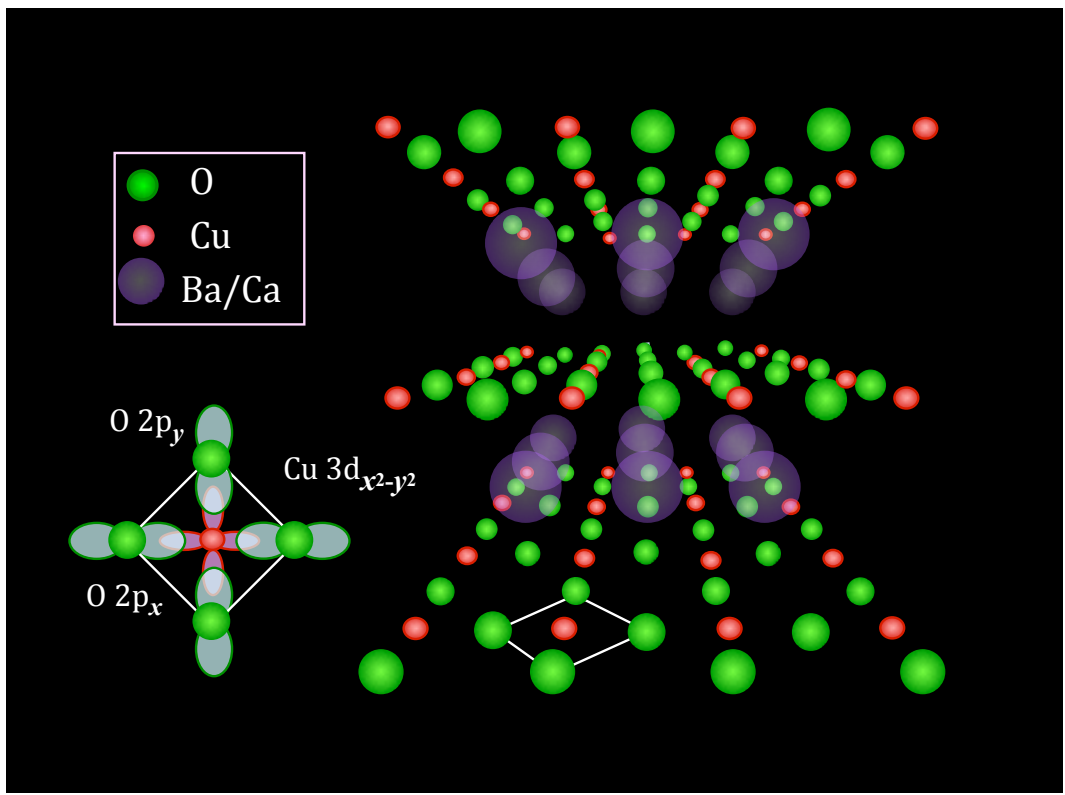

Figure 3 – **Crystal structure.** Layered cuprates are composed of CuO$_2$ planes typically separated by insulating spacer layers. The electronic structure of these planes primarily involves hybridization of a 3d $x^2$-$y^2$ hole on the copper sites with planar coordinated 2p$_x$ and 2p$_y$ oxygen orbitals.

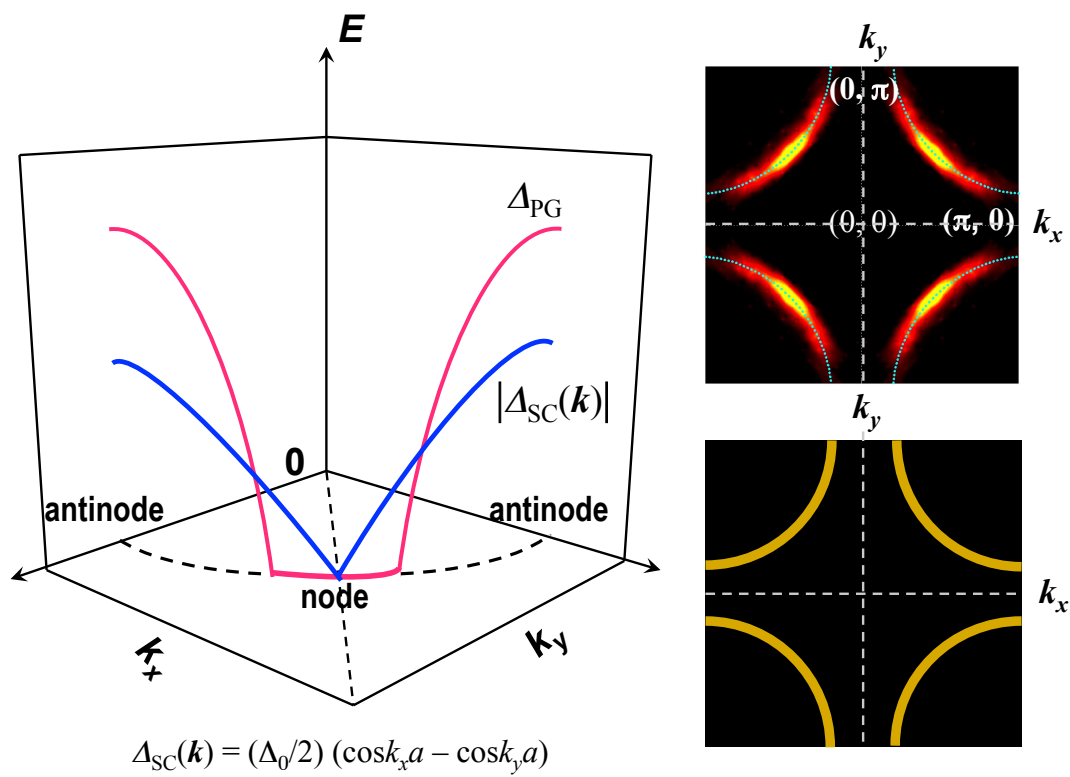

$\Delta_{SC}(k) = (\Delta_0/2)(\cos k_x a - \cos k_y a)$

Figure 4 – **Fermi surface, Fermi arcs, and gap functions.** The large Fermi surface predicted by band theory is observed by ARPES and STS for overdoped compounds (bottom right). But once the pseudogap sets in, the antinodal regions of the Fermi surface are gapped out, giving rise to Fermi arcs (top right). This is reflected (left) in the angle dependence of the superconducting gap (SC) and pseudogap (PG) around the underlying large Fermi surface (dashed curve) as revealed by ARPES and STS. Note the gapless region around the d-wave superconducting node for the PG case that defines the Fermi arcs. These arcs appear to be reconstructed into electron pockets centered at (Q/2,Q/2) once charge order sets in, as revealed by quantum oscillation studies, where (Q,0) is the charge order wavevector.